\documentclass[11pt,twoside]{article}

\usepackage{asp2014}

\aspSuppressVolSlug
\resetcounters

\begin{document}

\title{Generating background-corrected VPHAS+ HiPS with Montage}

\author{Thomas~Boch,$^1$ Caroline~Bot,$^1$ and Pierre~Fernique$^1$}
\affil{$^1$Observatoire astronomique de Strasbourg, Universit\'{e} de Strasbourg, CNRS, UMR 7550, Strasbourg, F-67000, France \email{thomas.boch@astro.unistra.fr}}

\paperauthor{Thomas~Boch}{thomas.boch@astro.unistra.fr}{orcid.org/0000-0001-5818-2781}{Universit\'e de Strasbourg, CNRS, UMR 7550}{Observatoire astronomique de Strasbourg}{Strasbourg}{}{67000}{France}
\paperauthor{Caroline Bot}{caroline.bot@astro.unistra.fr}{0000-0001-6118-2985}{Observatoire Astronomique de Strasbourg}{Centre de Donnees de Strasbourg}{Strasbourg}{}{67000}{France} 
\paperauthor{Pierre Fernique}{pierre.fernique@astro.unistra.fr}{0000-0002-3304-2923}{Observatoire Astronomique de Strasbourg}{Centre de Donnees de Strasbourg}{Strasbourg}{}{67000}{France} 



\begin{abstract}
This paper presents a technique for creating background-corrected HiPS (Hierarchical Progressive Surveys) from VPHAS+ images using the Montage toolkit. By combining advanced background correction methods and HiPS generation workflows, we produced high-quality color HiPS from the VST Photometric H-alpha Survey of the Southern Galactic Plane and Bulge (VPHAS+). These HiPS have been made publicly available through the HiPS network and serve as an invaluable addition to the HiPS ecosystem for astronomers worldwide.
\end{abstract}



\section{Introduction}

The VPHAS+ dataset is a comprehensive photometric survey conducted by the VLT Survey Telescope (VST). It maps the southern Galactic Plane and Bulge in multiple bands, including u, g, r, i, and H-alpha, at approximately 1 arcsecond angular resolution. The survey spans Galactic latitudes of -5° < b < +5° across all longitudes south of the celestial equator, with extensions reaching Galactic latitudes of ±10° near the Galactic Center.

\begin{figure}[!ht]
\centering
\includegraphics[width=1.0\textwidth]{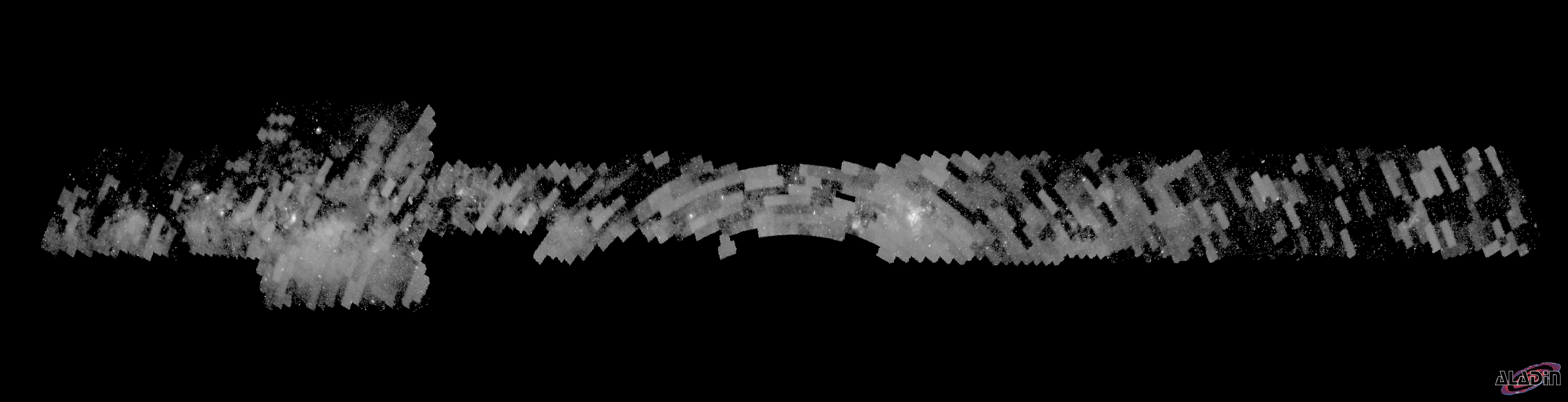}
    \caption{VPHAS+ g band HiPS, without background correction}
\label{fig:nocorr}
\end{figure}

HiPS \citep{2015A&A...578A.114F}, a widely adopted framework for multi-resolution visualization of large astronomical datasets, is well-suited to organize and display the VPHAS+ survey. However, HiPS generation of this survey requires careful background correction to address inherent variations in imaging data.\\
Figure~\ref{fig:nocorr} shows how the HiPS, generated by Hipsgen \citep{2022ASPC..532..467B}, looks when no background correction is performed.\\
This work describes our approach to overcoming these challenges using the Montage \citep{2017ASPC..512...81B} toolkit and highlights our results, including a color HiPS representation of VPHAS+ data.

\section{Methodology}
\subsection{Super-Tile construction}

\begin{figure}[!ht]
\centering
\includegraphics[width=1.0\textwidth]{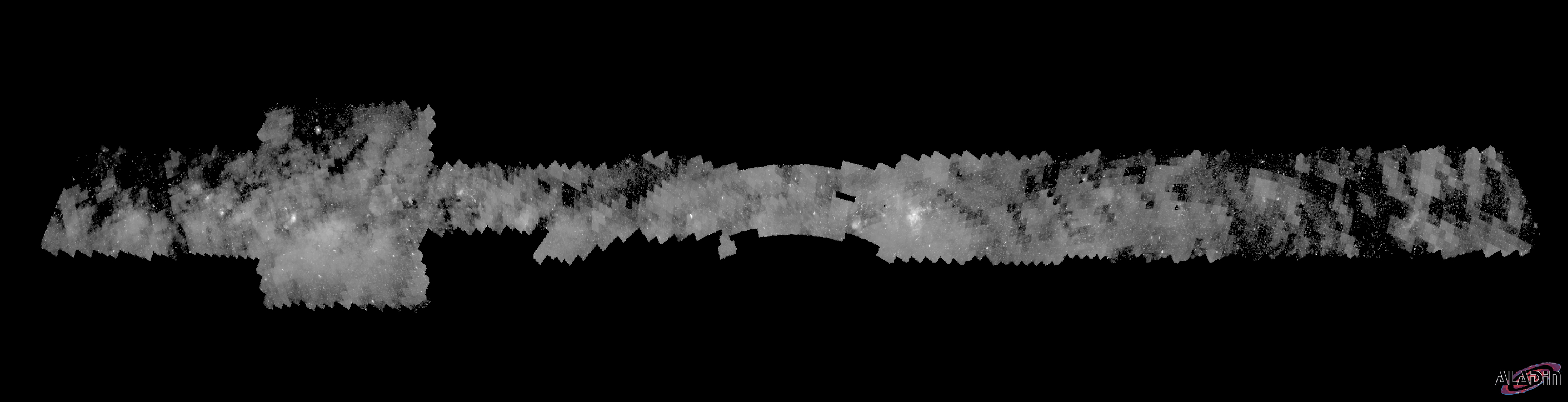}
    \caption{VPHAS+ g band HiPS, with background correction within each super-tile}
\label{fig:corr-supertile}
\end{figure}

For performance reasons, we organized the VPHAS+ data into 2700 super-tiles. Each super-tile consists of 32×32 regular HiPS tiles, with dimensions of 512×512 pixels per tile. To ensure seamless integration, we introduced a 512-pixel overlap between adjacent super-tiles. The super-tiles were processed independently, achieving homogeneous backgrounds within each, as shown in figure~\ref{fig:corr-supertile}.

\subsection{Background correction}

To rectify background variations between super-tiles, we performed the following steps:

\begin{itemize}
    \item Difference Computation: For each pair of overlapping super-tiles, differences were calculated using mDiffExec.
    \item Model Generation: Differences were modeled using mFitExec.
    \item Correction Application: Individual corrections were applied with mBgExec.
\end{itemize}

\begin{figure}[!ht]
\centering
\includegraphics[width=1.0\textwidth]{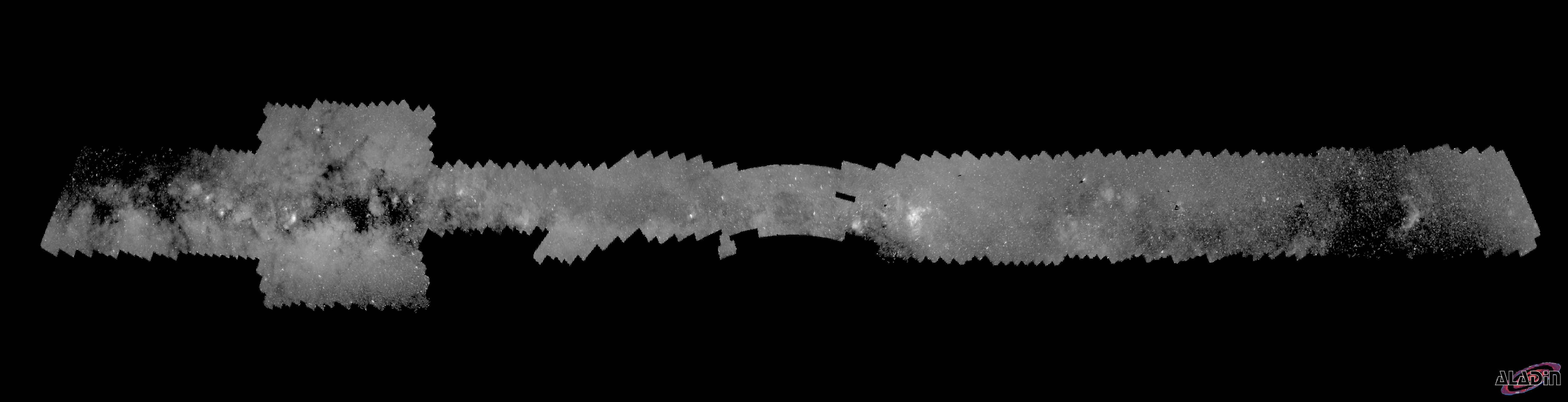}
    \caption{VPHAS+ g band HiPS, with all background corrections applied}
\label{fig:corr-all}
\end{figure}

The resulting global correction, shown in figure~\ref{fig:corr-all}, ensures a consistent background across the entire dataset.

\subsection{Regular tiles extraction}

Using a custom Python script, we extracted 1024 regular HiPS FITS tiles (at order 11) from each super-tile.

\subsection{Hierarchical HiPS construction}

The hierarchical HiPS structure was generated with \textit{Hipsgen}, starting from order 11 and progressively creating coarser-resolution layers.

\subsection{PNG tiles generation}

PNG tiles have been generated with mViewer,
taking advantage of the gaussian-log stretch.

\subsection{Color HiPS}

\begin{figure}[!ht]
\centering
\includegraphics[width=1.0\textwidth]{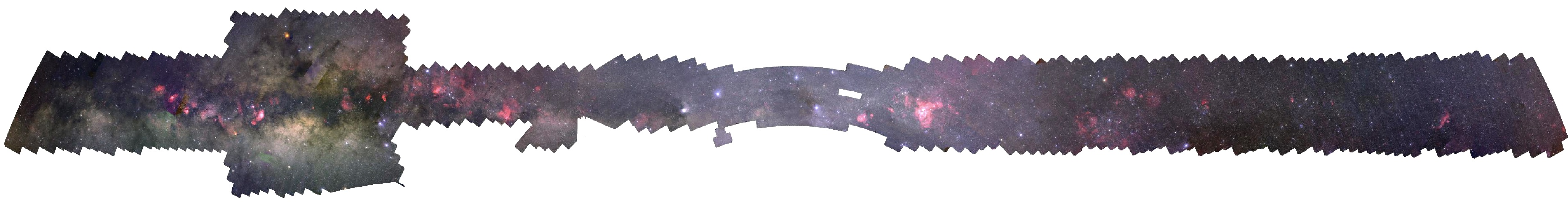}
    \caption{VPHAS+ color HiPS}
\label{fig:color-hips}
\end{figure}

For each of the 5 available bands (H-alpha, g, r, i, and u), we applied the HiPS creation process previously described. They are combined in a color HiPS, using this algorithm:

\begin{verbatim}
# first step
red   = (0.5*i + 0.33*u) / 0.83
green = r
blue = (g + 0.66*u) / 1.66

# second step
mask = h_alpha > threshold
red[mask] = np.maximum(1.2 * h_alpha[mask], red[mask])
\end{verbatim}
~~
The resulting HiPS is shown in figure~\ref{fig:color-hips}.

\section{Results}

The resulting VPHAS+ HiPS products show seamless background correction across all tiles and bands. The color HiPS provides a visually striking representation of the southern Galactic Plane and Bulge, enabling users to explore the survey interactively.

\subsection{HiPS Network Integration}

The VPHAS+ HiPS are now part of the HiPS network and accessible to the astronomical community. They represent a valuable resource for multi-wavelength studies of the Milky Way.\\
The page \verb|https://alasky.cds.unistra.fr/VPHAS/|, based on Aladin Lite \citep{2014ASPC..485..277B}, allows easy visualization of the HiPS generated.

\subsection{Applications}

This technique is not limited to VPHAS+ but can be extended to other astronomical datasets. We have already begun testing it for the IPHAS survey of the northern Galactic Plane.

\section{Conclusion}

We demonstrated a robust methodology for generating background-corrected HiPS from VPHAS+ images using Montage. This process addresses the challenges of background variability while preserving the scientific integrity of the data and enabling visualisation at all scales. The VPHAS+ HiPS and color HiPS products provide an important tool for astronomers studying the Galactic Plane, with potential applications across various datasets.

\acknowledgements We thank John Good for his valuable advice on Montage and background correction techniques.

\bibliography{P919}  

\begin{thebibliography}{}
\expandafter\ifx\csname natexlab\endcsname\relax\def\natexlab#1{#1}\fi
\expandafter\ifx\csname url\endcsname\relax
  \def\url#1{\texttt{#1}}\fi
\expandafter\ifx\csname urlprefix\endcsname\relax\def\urlprefix{URL }\fi
\providecommand{\eprint}[2][]{\url{#2}}

\bibitem[{{Berriman} et~al.(2017){Berriman}, {Good}, {Rusholme}, \&
  {Robitaille}}]{2017ASPC..512...81B}
{Berriman}, G.~B., {Good}, J.~C., {Rusholme}, B., \& {Robitaille}, T. 2017, in
  Astronomical Data Analysis Software and Systems XXV, edited by N.~P.~F.
  {Lorente}, K.~{Shortridge}, \& R.~{Wayth}, vol. 512 of Astronomical Society
  of the Pacific Conference Series, 81. \eprint{1608.02649}

\bibitem[{{Boch} \& {Fernique}(2014)}]{2014ASPC..485..277B}
{Boch}, T., \& {Fernique}, P. 2014, in Astronomical Data Analysis Software and
  Systems XXIII, edited by N.~{Manset}, \& P.~{Forshay}, vol. 485 of
  Astronomical Society of the Pacific Conference Series, 277

\bibitem[{{Bot} et~al.(2022){Bot}, {Fernique}, {Oberto}, {Durand}, {Boch},
  {Allen}, {Buga}, \& {Bonnarel}}]{2022ASPC..532..467B}
{Bot}, C., {Fernique}, P., {Oberto}, A., {Durand}, D., {Boch}, T., {Allen}, M.,
  {Buga}, M., \& {Bonnarel}, F. 2022, in Astronomical Data Analysis Software
  and Systems XXX, edited by J.~E. {Ruiz}, F.~{Pierfedereci}, \& P.~{Teuben},
  vol. 532 of Astronomical Society of the Pacific Conference Series, 467

\bibitem[{{Fernique} et~al.(2015){Fernique}, {Allen}, {Boch}, {Oberto},
  {Pineau}, {Durand}, {Bot}, {Cambr{\'e}sy}, {Derriere}, {Genova}, \&
  {Bonnarel}}]{2015A&A...578A.114F}
{Fernique}, P., {Allen}, M.~G., {Boch}, T., {Oberto}, A., {Pineau}, F.~X.,
  {Durand}, D., {Bot}, C., {Cambr{\'e}sy}, L., {Derriere}, S., {Genova}, F., \&
  {Bonnarel}, F. 2015, \aap, 578, A114. \eprint{1505.02291}

\end{thebibliography}


\end{document}